\documentclass[10pt, conference, compsocconf]{IEEEtran}
\IEEEoverridecommandlockouts
\usepackage{cite}
\usepackage{amsmath,amssymb,amsfonts}
\usepackage{graphicx}
\usepackage{textcomp}
\usepackage{xcolor}
\def\BibTeX{{\rm B\kern-.05em{\sc i\kern-.025em b}\kern-.08em
    T\kern-.1667em\lower.7ex\hbox{E}\kern-.125emX}}
\begin{document}

\title{Influence of A-Posteriori Subcell Limiting on Fault Frequency in Higher-Order DG Schemes*
\thanks{\textsuperscript{*} The underlying project has received funding from the European Union’s Horizon 2020 research and innovation  programme  under  grant  agreement  No  671698  (ExaHyPE).  All  software  is  freely  available  from www.exahype.eu.}
}

\author{%
\IEEEauthorblockN{1\textsuperscript{st} Anne Reinarz}
\IEEEauthorblockA{\textit{Department of Informatics} \\
\textit{Technical University of Munich}\\
Garching, Germany \\
reinarz@in.tum.de}
\and
\IEEEauthorblockN{2\textsuperscript{nd} Jean-Matthieu Gallard}
\IEEEauthorblockA{\textit{Department of Informatics} \\
\textit{Technical University of Munich}\\
Garching, Germany \\
gallard@in.tum.de}
\and
\IEEEauthorblockN{3\textsuperscript{rd} Michael Bader}
\IEEEauthorblockA{\textit{Department of Informatics} \\
\textit{Technical University of Munich}\\
Garching, Germany \\
bader@in.tum.de}
}

\maketitle

\begin{abstract}
Soft error rates are increasing as modern architectures require increasingly small features at low voltages.
Due to the large number of components used in HPC architectures, these are particularly vulnerable to soft errors. Hence, when designing applications that run for long time periods on large machines, algorithmic resilience must be taken into account. 
In this paper we analyse the inherent resiliency of a-posteriori limiting procedures in the context of the explicit ADER DG hyperbolic PDE solver ExaHyPE. 
The a-posteriori limiter checks element-local high-order DG solutions for physical admissibility, and can thus be expected to also detect hardware-induced errors.   
Algorithmically, it can be interpreted as element-local checkpointing and restarting of the solver with a more robust finite volume scheme on a fine subgrid.
We show that the limiter indeed increases the resilience of the DG algorithm, detecting and correcting particularly those faults which would otherwise lead to a fatal failure.
\end{abstract}


\section{Introduction}
Research on resilience often focuses on process failures such as the crash of a resource. These failures can be mitigated through process replication or checkpoint/restart procedures \cite{Herault:2015}. Checkpoint/restart techniques usually have high IO demands and can only be applied if the fault is detectable. However, faults which corrupt a computing system's state without affecting its functionality, referred to as soft or silent error are becoming increasing common. 
In future HPC architectures soft errors are expected to become even more frequent \cite{Argonne:2014} -- at exascale error rates are expected to force the entire software stack, including applications, to exploit any resilience-improving measures available. 

Soft errors occur, e.g., when particles such as neutrons from cosmic rays or alpha particles from packaging material reverse the data state of a memory cell \cite{Baumann:2005}. 
Failure rates are typically measured in FIT (failure in time), where one FIT corresponds to one failure every $10^9$ hours. DRAM soft error rates have remained relatively stable recently while SRAM soft error rates have been growing as memory chips become larger \cite{Mukherjee:2005}. Typical memory chips have $1000-10,000$ FIT/Mb \cite{Bronevetsky:2008}. For example, the ASCI Q experiences approximately $26$ CPU failures a week. This consitutes a higher than average error rate. It is due to the systems high altitude of about $2300$ m, where cosmic-ray-induced neutrons are roughly $6.4$ times more prevalent than at sea level \cite{Michalak:2005}. We perform our experiments on SuperMUC Phase 2 which consists of $2.6$ GHz $14$ core Intel Xeon E5-v3 processors each with a $72$Mb cache. Assuming a typical $2000$ FIT/Mb, running on $24$ nodes ($336$ cores) of SuperMUC for one day we would expect slightly more than $1$ error per day to occur. Assuming a higher, but still typical rate of $3000$ FIT/Mb, running on all $3072$ available nodes of superMUC phase 2, we expect over $400$ errors per day to occur, in other words, one error every 4 minutes. Hence, even though errors are rare events in terms of numerical kernels (which might be called billions of times before encountering a soft error), error frequencies in the range of hours or even minutes can strongly impede the efficiency of checkpoint-restart approaches \cite{Varela:2010}. 

While there is research into improving chip design to reduce soft error rates (SER), see e.g. \cite{Mitra:2005}, in this paper we focus on the design of resilient algorithms.
Several groups have studied the influence of soft errors on various  iterative linear methods, largely because so many scientific applications rely on them, see e.g. \cite{Bronevetsky:2008, Bronevetsky:2018, Austin:2015, Shantharam:2012}. In this paper we examine a complex application consisting of an explicit PDE solver.
The solver adopts a high-order discontinuous Galerkin (DG) scheme which requires a limiting procedure to cope with oscillations that may occur at strong gradients or even discontinuities in the solution. 
Here, we adopt an \emph{a-posteriori} approach that first computes element-local high-order DG solution, which are then checked for physical and numerical admissibility. 
Solution candidates for which the numerical scheme fails, are disregarded and recomputed \cite{Loubere:2013}. 
We aim to exploit this compute--check--recompute procedure also for faulty solutions resulting from soft errors.    

The numerical solver is developed as part of the ExaHyPE project (www.exahype.eu), which focuses on the implementation of an explicit solver for systems of hyperbolic PDEs. Hyperbolic PDE systems describe many important physical phenomena, such as earthquake and tsunami waves, or even  gravitational waves.  The two primary applications currently tackled by the ExaHyPE engine are regional seismic risk assessment  and the simulation of gravitational waves emitted by binary neutron stars, see e.g. \cite{Dumbser:2017,  Bishop:2016, Tsorkaros:2016, Duru:2017, Rannabauer:2018}.
Both of these applications require the simulation of large times scales and large  domains to compute new physically relevant results. Current production runs of the seismology application require almost two days on approximately $800$ cores to complete. However, we expect future scenarios to be considerably larger. This makes detection and correction of soft errors important. 

The ExaHyPE engine relies on the adaptive mesh refinement (AMR) functionality provided by the Peano dynamic AMR framework \cite{Weinzierl:2011}. Peano provides data structures and so-called hooks for the numerical kernels, which are used to implement the higher-order ADER DG scheme, see e.g. \cite{Titarev:2002, Zanotti:2015, Charrier:2018}. To handle shocks and discontinuities an a-posteriori subcell limiter \cite{Loubere:2013} is applied. 
The Peano framework allows an element-local check-and-recompute logic to deal with ``troubled'' elements, i.e. those elements for which certain physical and numerical admissibility criteria are  not met. In the troubled elements the solution is recomputed with a robust finite volume scheme.

In the following we start with a brief overview of the problem setting and the solution methods used in the ExaHyPE Engine. We particularly focus on the limiting procedure and how the limiter provides intrinsic resiliency properties of the entire numerical algorithm. Next we discuss the methodology used to insert soft errors, and give some numerical results. We then discuss the role of the limiter in detecting certain types of soft errors and analyse a restart procedure used when a soft fault occurs during the limiter phase.

\section{Problem Setting}\label{sec:scenarios}
The ExaHyPE engine can solve a large class of systems of first-order hyperbolic PDEs. In this paper we focus on equations of the form:
\begin{quote}
Find a space- and time-dependent state vector $\textbf{Q}(x,t) \subset \mathbb{R}^q$ such that  
\begin{equation} \label{eq:hyperbolicPDE}
\frac{\partial \mathbf{Q}}{\partial t}(x,t) + \nabla \cdot \mathbf{F}(\mathbf{Q})
+ \mathbf{B}(\mathbf{Q})\cdot \nabla\mathbf{Q}(x,t) = \mathbf{S}(\mathbf{Q})
\end{equation}
for any $x\in \Omega \subset \mathbb{R}^d$ and $t\in\mathbb{R}_0^+$. 
\end{quote} 
Here, $\mathbf{F}$ denotes the conserved flux vector, $\mathbf{B}$ denotes the nonconservative (system) matrix composing the non-conservative fluxes and $\mathbf{S}(\mathbf{Q})$ denotes the source terms.

The equation is supplemented by appropriate initial conditions
\begin{equation}\label{eq:initial}
 \mathbf{Q}(x,t=0)=\mathbf{Q}_0(x),~\forall x\in \Omega,
\end{equation}
and boundary conditions
\begin{equation}\label{eq:boundary}
\mathbf{Q}(x,t)=\mathbf{Q}_B(x,t),~~\forall x\in \partial\Omega,~t\in\mathbb{R}_0^+,
\end{equation}

To test the resilience of the algorithm we use two very different scenarios, the Euler equations and the Einstein equations.

\subsection{The Euler Equations}
The Euler equations model the flow of an inviscid fluid with constant density.
Solutions of the Euler equations are often used as approximations to real (i.e., viscous) fluids problem,
e.g.\ the lift of a thin airfoil.

As a non-linear test we focus on the Euler equations of compressible gas dynamics
\begin{equation}\label{eq:euler}
\frac{\partial}{\partial t} \textbf{Q}
+
\nabla\cdot\textbf{F}(\textbf{Q})
= 0,
\end{equation}
where
\begin{equation}\label{eq:euler2}
\textbf{Q} = \begin{pmatrix}
\rho\\\textbf{j}\\\ E
\end{pmatrix}
\quad \mbox{and} \quad
\textbf{F(Q)} = 
\begin{pmatrix}
\textbf{j}\\
\frac{1}{\rho}\textbf{j}\otimes\textbf{j} + p I \\
\frac{1}{\rho}\textbf{j}\,(E + p)
\end{pmatrix},
\end{equation}
where $\rho$ denotes the mass density, $\bf{j}\in\mathbb{R}^d$ denotes the momentum density, $E$ denotes the energy density. Further, $p$ denotes the fluid pressure, which given by an equation of state. Here, the perfect gas law with adiabatic index $\gamma$ is used:
\begin{equation}
 p=(\gamma-1)\left( E-\frac{1}{2\rho}\textbf{j}^2\right).
\end{equation}
We test two benchmark scenarios for the Euler equations. The first is a smooth test case consisting of a  moving Gaussian matter distribution and the second is the Sod's shock tube problem in 2D \cite{Toro:2009}. For both of these test cases analytical solutions are available to allow us to easily measure errors in the simulation.

\subsection{The Einstein Equations}
The Einstein equations describe the theory of gravitation given by the equality of mass, energy and curved spacetime. When enough mass is compressed in a small area to form a black hole the classical Newtonian theory of gravity breaks down and can only be described using the Einstein equations. The spacetime of a black hole outside of the black hole center point is approximately given by a vacuum solution, i.e.
\begin{equation}
G_{\mu\nu} \sim G M \delta(x),
\end{equation}
where $G_{\mu\nu}$ is the Einstein tensor. It has a zero source when the Dirac distribution $\delta$ gives no contribution, i.e. at $x>0$.

The formulation of Einstein's equations as an initial value problem suitable for numerical integration is challenging, within ExaHyPE a first order strongly hyperbolic formulation of the constraint damping Z4 formulation of Einstein's equations is implemented \cite{Dumbser:2017}. In this first-order formulation the Einstein equations can be written in the form: 
\begin{equation}\label{eq:einstein}
\frac{\partial \mathbf{Q}}{\partial t}
 +   \mathbf{B} (\mathbf{Q}) \cdot \nabla \mathbf{Q} = \mathbf{S}(\mathbf{Q}).
\end{equation}
For brevity we do not give the form of the non-conservative system matrix $\mathbf{B}$, or of the state vector $\mathbf{Q}\subset\mathbb{R}^{59}$ in their entirety, the formulation is described in detail by Dumbser et. al. \cite{Dumbser:2017}.
In this formulation, there are $59$ scalar variables which describe the Einstein tensor.

We test two simple benchmark scenarios for the Einstein equations. The first is the so called gauge wave: here, gauge conditions are introduced in a vacuum and the metric takes the form of a sine wave. The gauge wave test evolves flat empty space and the sine wave simply propagates. This gives a smooth solution \cite{Rezzolla:2013}. The second is the evolution in time of a single static black hole. This test requires the use of a limiter \cite{Rezzolla:2013}.

\section{The ADER-DG Algorithm}
In this section we provide a brief overview of the method used to solve the system (\ref{eq:hyperbolicPDE}). The arbitrary high-order accurate ADER Discontinuous Galerkin (DG) method was introduced by Toro and Titarev in 2002 \cite{Titarev:2002}. 
In ExaHyPE we use the formulation developed by Dumbser et al. \cite{Zanotti:2015}.
The solver phases have been reordered to allow a more memory- and communication-efficient implementation of the method \cite{ Charrier:2018}. 

The computational domain $\Omega\subset \mathbb{R}^d$ with $d=2$ or $3$  is discretised on a space-time adaptive cartesian  grid $\Omega=\bigcup_{i} T_i$. 
%
At the beginning of each time step, the numerical solution of (\ref{eq:hyperbolicPDE}) for the state vector $\mathbf{Q}$ is represented within each mesh element by:
 \begin{equation}\label{eq:DGAnsatz}
  u_h(x,t^n) = \sum_\ell \phi_\ell(x)\hat u_\ell^n,
 \end{equation}
where $\phi_k$ are basis functions from the space of piecewise polynomias of degree $N$ constructed using tensor products of Lagrange polynomials over Gauss-Legendre or Gauss-Lobatto points. In DG methods the basis functions have compact support on each cell. 

Then the DG ansatz function $u_h$ is inserted into equation (\ref{eq:hyperbolicPDE}) and multiplied with a test function $\phi_k$ and integrated over a space-time control volume $T_i \times [t^n ,t^{n+1} ]$ to get the weak form of the equation.  This is the so-called predictor step, a purely element-local calculation that neglects the impact of neighbouring cells.
 
The arising jump terms on cell interfaces are subjected to a Riemann solver, denoted by $\mathcal{D^-}$, as its used in Godunov-type finite volume (FV) schemes. The non-conservative product on the element boundaries is discretized via a path-conservative jump term. These solution of the Riemann problems on each element is embarrassingly parallel and typically low in arithmetic intensity. Next we introduce an element-local space-time predictor $q_h(x,t)$ with left and right states $q_h^-$ and $q_h^+$, which combines the initial predictor and the solution of the Riemann problem. Thus, the fully discrete one-step ADER-DG scheme reads:
\begin{displaymath}
\begin{split}
\Big(\int_{T_i}&\phi_k\phi_ldx\Big)(\hat u_l^{n+1}-\hat u_l^n) \\&+ \int_{t^n}^{t^{n+1}}\int_{\partial T_i} \phi_k \mathcal{D^-}(q_h^-,q_h^+)\cdot n \, dS\,dt\\
                &+ \int_{t^n}^{t^{n+1}}\int_{T_i\backslash\partial T_i} \phi_k\left(\nabla\cdot \textbf{F}(q_h)+ \textbf{B}(q_h)\cdot\nabla q_h\right)\, dx\,dt\\
                &= \int_{t^n}^{t^{n+1}}\int_{T_i} \phi_k S(q_h)\,dx\,dt.
\end{split}
\end{displaymath}

 The space time predictor $q_h$ is defined in terms of a set of nodal space time basis functions 
 as the solution to an element-local system that can be solved with a fast converging discrete Picard iteration.

Before moving to the next time step, the Courant-Friedrichs-Lewy (CFL) number needs to be calculated. This number gives a restriction  on the size of the next time step. 

Note that this unlimited ADER DG version will suffer from numerical oscillations (Gibbs phenomenon) in the presence of steep gradients or shock waves. Therefore a limiter must be applied. We describe it in more detail in the following section, as it is relevant to the resilience of the algorithm.

\subsection{Limiting Procedure}\label{sec:limiter}

The ExaHyPE Engine uses the multi-dimensional optimal-order-detection (MOOD) approach. This limiter was initially applied to finite volume schemes and has recently been extended to DG schemes \cite{Loubere:2013}. The MOOD approach bypasses issues with a-priori detection of troubled zones and allows for good resolution of shocks and other discontinuities. The approach preserves the subcell resolution properties of the ADER DG algorithm and has no parameters that need to be tuned.

\begin{figure}[t]
  \centerline{\includegraphics[width=0.45\textwidth]{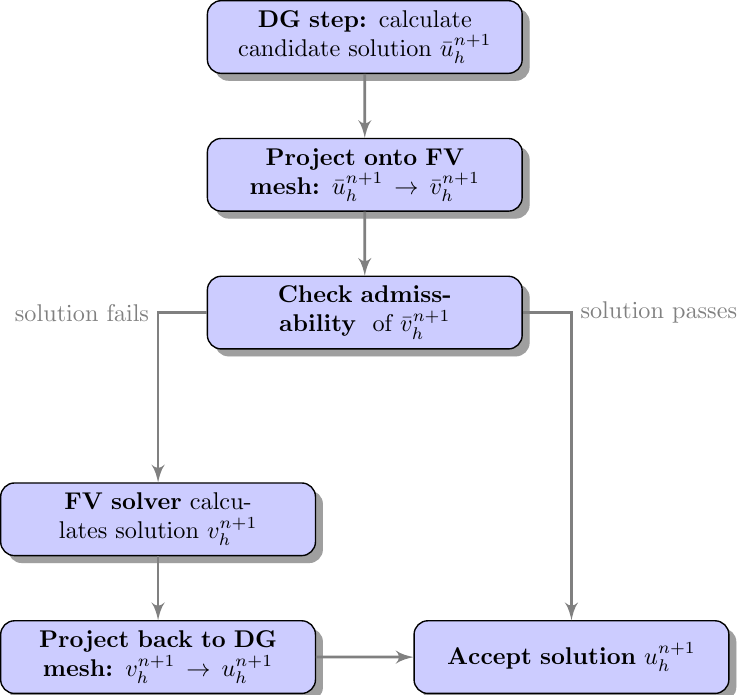}}
 \caption{Schematic of the a-posteriori limiting approach for the time step $n+1$. }
 \label{fig:flowchart}
\end{figure}

In this approach the solution is checked a-posteriori for certain admissibility criteria and is recalculated with a more robust numerical scheme if it does not meet them. As very simple a-posteriori detection criteria we use
\begin{enumerate}
 \item \textit{Physical Admissability Criteria}: Depending on the system of PDEs being studied certain physical constraints can be place on the solution. For the Euler equations these are positivity of the density and  energy.
 \item \textit{Numerical Admissability Criteria}: We use two numerical admissability criteria, the first is an absence of floating point errors, i.e. NaNs. This criterion can be checked cheaply. The second is a relaxed discrete maximum principle (DMP) in the sense of polynomials.
 \end{enumerate}
A candidate solution $u^*_h(x,t^{n+1})$ is said to fulfill the DMP in cell $T_i$ if the following relation is fulfilled componentwise for all conserved variables and for all points $x\in T_i$:
\begin{equation}\label{eq:DMP}
 \min_{y\in V_i} u_h(y,t^n)-\delta \leq u_h^*(x,t^{n+1}) \leq u_h(y,t^n) + \delta,
\end{equation}
where $V_i$ is a set containing the element $T_i$ and its Voronoi neighbor cells that share a common node with $T_i$. The strict maximum principle is relaxed using $\delta$ in order to allow very small overshoots and undershoots. This approach avoids problems with roundoff errors that would occur when applying (\ref{eq:DMP}) in a strict way, i.e. with $\delta=0$.  This criterion can be very expensive to check, but is necessary in order to detect shocks and discontinuities.

If one of these criteria is violated after a time step, the scheme goes back to the old time step and recomputes the solution in the troubled cells, using a more robust high resolution shock capturing FV scheme on a fine subgrid composed of $(2N+1)^d$ subcells. A schematic of the limiting algorithm is shown in Figure \ref{fig:flowchart}. Note that this is embedded in the ADER DG scheme discussed in the previous section.

The limiter can be interpreted as element-local checkpointing and restarting of the solver with a more robust scheme on a fine subgrid. We claim that in many cases also soft errors can be detected by means of the a-posteriori detection criterion. For example, in the case of a sign change, the physical admissibility constraint will trigger a recalculation in that element. Further, the DMP can detect changes of the state vector that have a large magnitude. This means that particularly those changes which could lead to a large error  or numerical instabilities are filtered out by the limiting procedure.

\section{Identification of compute-intensive kernels}

In Table \ref{table:time_in_kernel} we list the time spent in aggregated kernel time compared to the total execution time. We observe that overheads for mesh traversals, etc., in the engine and the underlying Peano framework are compensated well, once the number of quantities in the PDE  or the discretisation order become high enough.  
\begin{table}[b]
\caption{Time spent in the kernels expressed as a percent of the total runtime. }
\begin{center}
\begin{tabular}{|l|c|c|c|c|c|c|} \hline 
      &       \multicolumn{2}{c|}{}          &   \multicolumn{2}{c|}{}\\[-.8em]
 & \multicolumn{2}{c|}{Without Limiting} &\multicolumn{2}{c|}{Limiting}\\
      &       &          &       &\\[-.8em]
Order & Euler & Einstein & Euler & Einstein \\\hline
      &       &          &       &\\[-.8em]
3 & 39.0\%  & 98.6\% & 10.2\% & 59.6\%   \\
6 & 84.1\%  & 99.8\% & 38.7\% & 74.4\%  \\
9 & 94.9\%  & 99.7\% & 62.4\% & -  \\
\hline
\end{tabular}
\end{center}
\label{table:time_in_kernel}
\end{table}

In order to identify the most likely places for an error to occur we need to consider the time spent in each part of the application. 
We evaluated the performance on the benchmark scenarios described in Section \ref{sec:scenarios}, with and without activating the limiter. Without limiting the moving Gaussian test case is used for Euler equations and the Gauge wave test is used for the Einstein equations. With limiting, the sod shock tube test case is used for the Euler equations and the static black hole is used for the Einstein equations.

\subsection{Cost of ADER-DG}
In Table \ref{table:percentage_nolim}, we list the percentage of execution time spent in each of the ExaHyPE kernel functions for increasing discretisation order ($3$, $6$ and $9$) for the two scenarios without a limiter. Here, fSTPVI is the kernel used for space-time predictor and volume integral, often it is advantageous to fuse the three loops of ADER-DG into a single loop over the mesh \cite{Charrier:2018}. Further, Riemann is the kernel which solves the Riemann problem and CFL is the kernel which calculates the next stable time step. It immediately becomes clear that fSTPVI  is by far the most expensive kernel and dominates the execution time in all setups. Its dominance increases with higher orders and number of quantities. Thus, we focus on soft errors which occur during the calculation of this kernel.

\begin{table}[t]
\caption{Percentage of execution time spent in the various ExaHyPE kernels for two non-limited scenarios.}
\begin{center}
\begin{tabular}{|l|c|c|c|c|c|c|} \hline
&\multicolumn{3}{c|}{}&\multicolumn{3}{c|}{}\\[-.8em]
Kernel/ & \multicolumn{3}{c|}{Euler ($q=5$)} & \multicolumn{3}{c|}{Einstein ($q=59$)} \\
&&&&&&\\[-.8em]
Order & 3 & 6 & 9 & 3 & 6 & 9 \\\hline
&&&&&&\\[-.8em]
fSTPVI              & 73.5\% & 87.4\% & 91.9\%  & 97.7\% & 99.1\% & 99.4\% \\
Riemann             & 7.7\%  & 2.0\%  & 1.0\%   & 1.8\%  & 0.6\%  & 0.3\% \\
CFL                 & 7.5\%  & 4.7\%  & 3.3\%   & 0.1\%  & 0.1\%  & 0.0\% \\
other               & 11.3\% & 5.9\%  & 3.8\%   & 0.4\%  & 0.2\%  & 0.3\% \\
\hline
\end{tabular}
\end{center}
\label{table:percentage_nolim}
\end{table}

\subsection{Cost of limiting}\label{sec:cost_limiter}

\begin{table}[b]
\caption{Distribution of execution time spent in ExaHyPE's limiter kernels.  The last two rows compare the aggregate percentage of kernel time spent in limiting vs.\ basic ADER-DG.
} 
\begin{center}
\resizebox{\linewidth}{!}{%
\begin{tabular}{|p{1.5cm}|c|c|c|c|c|c|} \hline
Kernel/ & \multicolumn{3}{c|}{Euler ($q=5$)} & \multicolumn{3}{c|}{Einstein ($q=59$)} \\
Order & 3 & 6 & 9 & 3 & 6 & 9 \\\hline
project~to~FV & 7.7\% & 9.0\% & 11.0\% & 0.4\% & 0.6\% & 2.4\%  \\
project~to~DG & 16.0\% & 15.3\% & 17.8\% & 0.4\% & 1.1\%  & 7.0\%  \\
admissability          & 31.2\% & 49.5\% & 60.4\% & 3.5\% & 9.0\% & 14.4\% \\
\hline
&&&&&&\\[-.8em]
\textbf{aggregate limiter} & \textbf{55.1\%} & \textbf{73.9\%} & \textbf{89.2\%} & \textbf{4.2\%} & \textbf{10.6\%} & \textbf{23.8\%} \\
\hline
&&&&&&\\[-.8em]
aggregate ADER-DG & 44.9\% & 26.1\% & 10.8\% & 95.8\% & 89.4\% & 76.1\% \\\hline
\end{tabular}
}
\end{center}
\label{table:percentage_lim}
\end{table}

Next we evaluate the effect of the limiter on the run time. In Table \ref{table:percentage_lim}, admissability refers to the calculations needed to check for physical and numerical admissability as described in Section \ref{sec:limiter}, project to FV and DG refer to the time spent projecting to the FV mesh and back.  Table \ref{table:percentage_lim} lists how the execution time of kernels is distributed between kernels for limiting vs.\ the basic ADER-DG implementation, again for increasing discretisation order ($3$, $6$ and $9$). The time spent in the limiter is dominated by the time needed to calculate the discrete maximum principle and for increasing order the limiter starts to dominate the total solution time.

\section{Fault injection methodology}
 One approach to evaluating the effect of soft faults uses a numerical uncertainty fault model in order to ascertain worst case scenarios instead of average behaviour \cite{Elliott:2014}. However, by far the most common technique used to evaluate the effect of soft errors is the injection of faults at random points in the execution of the application \cite{Lu:2004}. This approach has the advantage of allowing the behaviour of extremely complex algorithms to be evaluated.

All tensors and matrices in the ADER-DG formulation are stored as regular arrays, here dimensions always include the quadrature points within an element and the quantities of the PDE. The choice of fastest-running index of the array corresponds to the conflict of array-of-struct (AoS) vs. struct-of-array (SoA) data layout, where a ``struct" typically refers to the quantities at one interpolation point. We model the impact of soft errors by flipping a single randomly chosen bit of the state vector.
We insert the soft error at the beginning of a randomly chosen time step. In particular, we do not flip bits in function return pointers. 

We first consider the following possible outcomes of a  bit flip (compare e.g.\ \cite{Pawelczak:2017}) on the run of the ADER-DG solver:
\begin{itemize}
 \item Success: the solver completes successfully with a negligible error compared to the reference solution,
 \item Silent Data Corruptions (SDC): the solver completes, but there is non-negligible error
 \item Detectable Correctable Errors (DCE)
 \item Detectable Uncorrectable Errors (DUE)
\end{itemize}
We compute the numerical error at a time step $T$ by:
\begin{equation}\label{eq:l2norm}
 \text{error} = \max_{i=0...q} \sqrt{\int_{\Omega} (\mathbf{Q}_i(x,T)- \mathbf{Q}^\text{ref}_i(x,T))^2 dx}.
\end{equation}
Here, the reference solution $\mathbf{Q}^\text{ref}$ is not the analytically known exact solution, but a reference calculated without introducing any bitflips.
We consider an error negligible if the relative error is less than $10^{-14}$. 

\section{Impact of Soft Errors}
All tests in this section were run on $2.6$ GHz $14$ core Intel Xeon E5-v3 processors (SuperMUC Phase~2). The tests were run on one node, consisting of 2 processors with 14 cores each using Intel's TBB \cite{Reinders:2007} for parallelisation.

\subsection{The Euler Equations}

To compare limited and unlimited runs, the experiment in this section uses the smooth initial conditions given in Section \ref{sec:scenarios}. The mesh consists of  $243$ elements in each direction with an increasing discretisation order of $3$, $6$ and $9$. These meshes consist of approximately $4\cdot 10^6$ for $N=3$, $10^7$ for $N=6$ and $3\cdot 10^7$ for $N=9$ degrees of freedom per time step. 
The run continued for $200$ time steps. The scenario was run $600$ times without a limiter and $600$ times with a limiter for each discretisation order. 

\begin{figure}[b]
  \centerline{\includegraphics[width=0.45\textwidth]{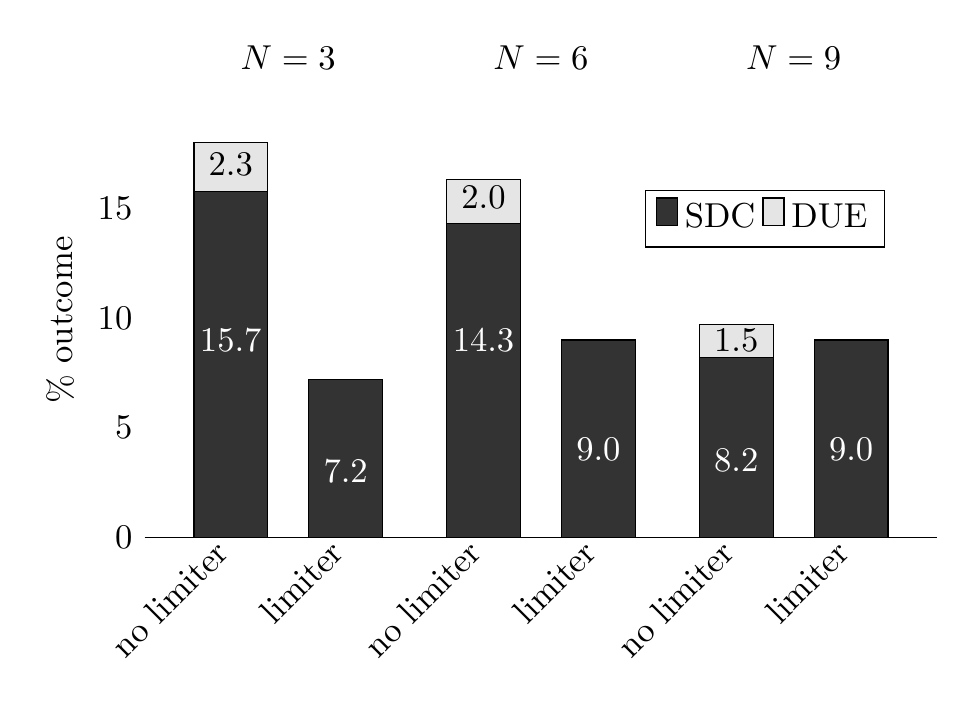}}
 \caption{Effects of a single  bitflip on the Euler test case. The $y$ axis gives the percentage of runs with the outcome listed on the $x$ axis. 
 }
 \label{fig:initialEuler}
\end{figure}

The outcomes of the runs for the three different discretisation orders are shown in Figure \ref{fig:initialEuler}. Of those test cases run without limiter for degrees $N=3$ and $6$ in over $80\%$ of the cases the error was negligible. In the case of order $N=9$ this number is at about $90\%$. The source of the discrepancy is unclear. In all test cases the largest measured error was only $10^{-6}$. However, we note that for all discretisation orders approximately $2\%$ of the injected bitflips led to the program aborting. When the limiter was applied the number of cases of silent data corruption (SDC) fell slightly for all discretisation orders. However, the most notable outcome of applying the limiter is that there were no longer any cases in which the program aborted.

\begin{figure}[t]
\begin{centering}
  \includegraphics[width=0.23\textwidth]{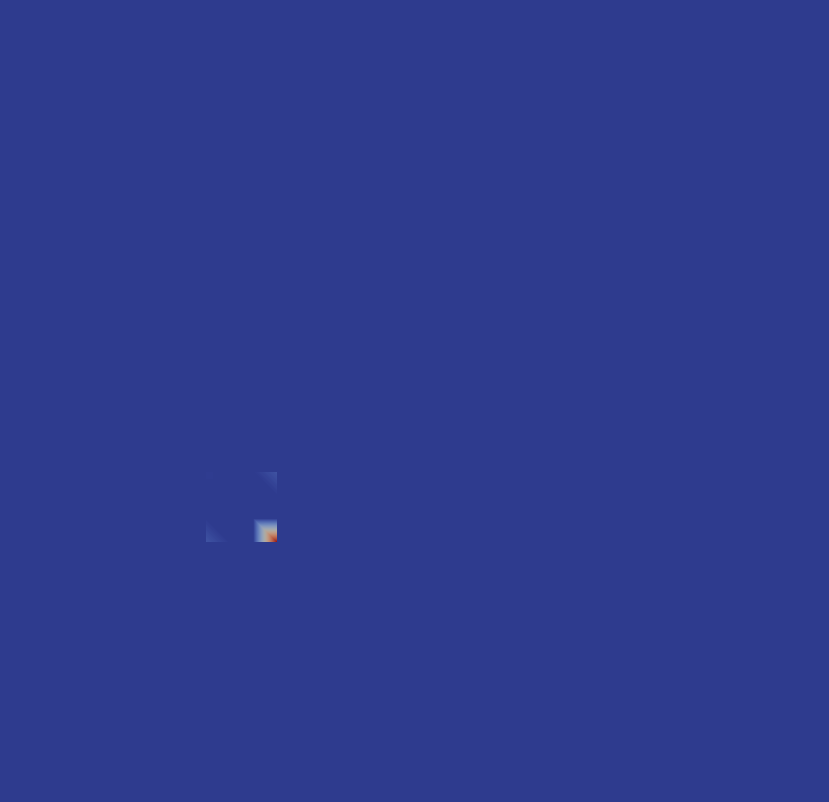}
  \includegraphics[width=0.23\textwidth]{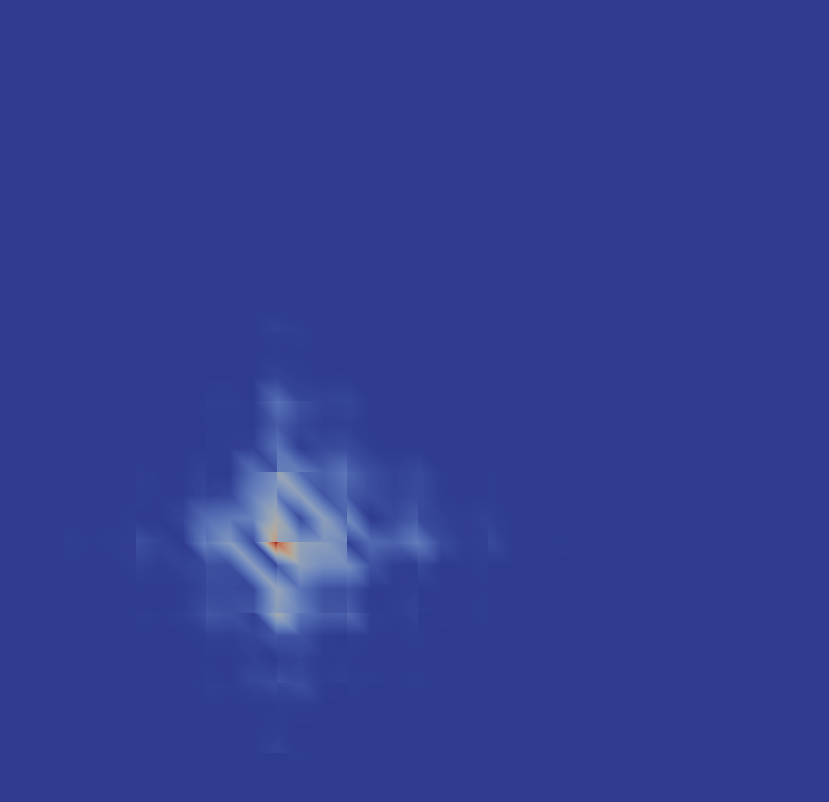}\\
  \vspace{4pt}
  \hspace{5.4pt}\includegraphics[width=0.23\textwidth]{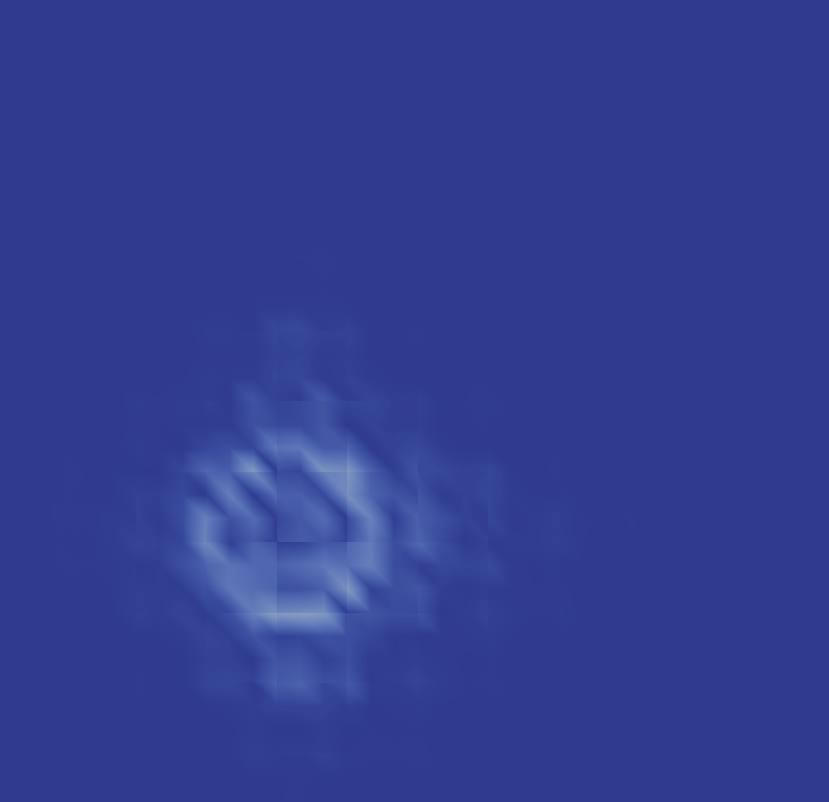}
  \includegraphics[width=0.23\textwidth]{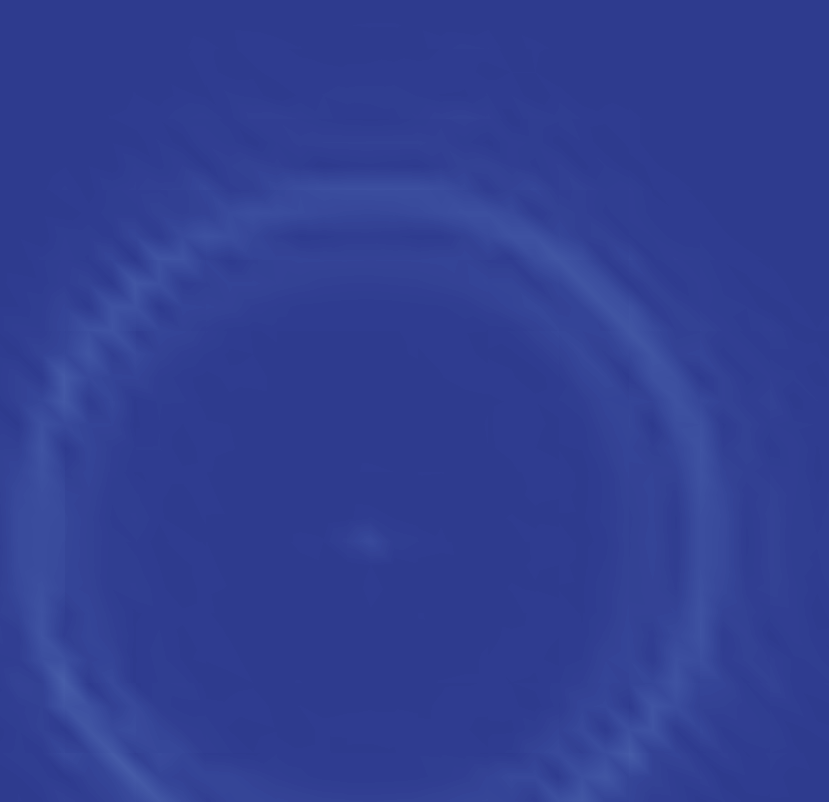}
 \end{centering}
 \caption{Pointwise error of the solution at 4 different time steps.}
 \label{fig:errorprop}
\end{figure}

 To visualise these results we chose the test in which the solution had a large error of $10^{-6}$ and plotted the pointwise error compared to an analytically calculated reference solution. The propagation of the error through several time steps is shown in Figure \ref{fig:errorprop}. The error which originates at only one point and in one variable propagates through the domain in the same manner as a point source introduced into the program normally would. 

\begin{table}[t]
\caption{Frequency of outcome and average time step at which the bitflip was inserted for different outcomes of the simulation without a limiter.
}
\begin{center}
\begin{tabular}{|c|c|c|c|c|c|c|}\hline
                & \multicolumn{6}{c|}{ Without limiting} \\
                & \multicolumn{2}{c|}{$N=3$} & \multicolumn{2}{c|}{$N=6$} & \multicolumn{2}{c|}{$N=9$} \\
Case            & outcome & time & outcome & time & outcome & time\\\hline
success         & 82.0\% & $118$ & 83.7\% & 101   & 90.3\% & 101\\
SDC             & 15.7\% & $81$  & 14.3\% & 66    & 8.2\%  & 87\\
DUE             & 2.3\%  & $31$  & 2.0\%  & 64    & 1.5\%  & 84\\
\hline
\end{tabular}
\end{center}
\label{table:timestep1}
\end{table}
\begin{table}[t]
\caption{Frequency of outcome and average time step at which the bitflip was inserted for different outcomes of the simulation with a limiter.
}
\begin{center}
\begin{tabular}{|c|c|c|c|c|c|c|}\hline
                & \multicolumn{6}{c|}{With limiting} \\
                & \multicolumn{2}{c|}{$N=3$} & \multicolumn{2}{c|}{$N=6$} & \multicolumn{2}{c|}{$N=9$} \\
Case            & outcome & time & outcome & time & outcome & time\\\hline
success         & 92.8\% & $106$ & 91.0\% & $102$   & 91.0\%  & $100$   \\
SDC           & 7.2\%  & $23$  & 9.0\%  & $81$    & 9.0\%  & $98$   \\
DUE           & 0\%    & -     & 0\%    & -       & 0\%     & -  \\
\hline
\end{tabular}
\end{center}
\label{table:timestep2}
\end{table}

Tables \ref{table:timestep1} and \ref{table:timestep2} shows the average time step at which a bitflip occured for the three different outcomes. The overall average time step at which a bitflip was introduced was approximately $100$ in all cases. Errors introduced earlier in the program were much more likely  to lead to a high error or to the program aborting in all setups except $N=9$.

To understand the cases in which the program aborted, we examined the types of bitflips leading to this outcome. They fell loosely into three categories: sign changes, change into a number close to zero and changes of very large magnitude.
In the case of sign changes, the program aborted only when the sign change led to a physically inadmissible state, such as a negative density or energy. Sign changes in the velocity, on the other hand, did not lead to the program aborting. The same was true of bitflips which resulted in numbers very close to $0$, this was only observed to lead to the program aborting if the change occured in the density. This behaviour is easily explained by the fact that a division by the density occurs in the Euler equations and a density of $0$ is thus not allowable. Bitflips which led to changes of an extremely large magnitude caused the program to abort independent of the variable in which they occured.

Since a change which might merely lead to a small error in one variable can lead to an early termination in another, we conclude that these results are at least somewhat application-specific and cannot readily be generalised to other systems of PDE. However, the capabalities of the limiter to detect and correct in particular those changes which lead to numerical instabilities should extend also to other applications as long as the physical admissibility criteria are set correctly. To see how generalisable the results are we next test a very different application: the Einstein equations. 

\subsection{The Einstein Equations}

\begin{table}[t]
\caption{Frequency of outcome and average time step at which the bitflip was inserted for different outcomes for the Einstein equations.
}
\begin{center}
\begin{tabular}{|c|c|c|c|c|}\hline
                & \multicolumn{2}{c|}{With limiting} & \multicolumn{2}{c|}{Limiting}\\
Case            & outcome & time & outcome & time \\\hline
success         & 4\%    & 100 & 2.7\%  & 100   \\
SDC      & 95.5\% & 100 & 97.3\%  & 100     \\
DUE           & 0.5\%  & 69  & 0\%  & -       \\
\hline
\end{tabular}
\end{center}
\label{table:timestep3}
\end{table}


The Einstein equations are a much larger system of PDEs, we move from $5$ variables to $59$.  To facillitate comparison between the limited and unlimited case we use the gauge wave test case. Due to this large size and since we want to be able to perform our tests on one node, we choose a smaller mesh of $27$ elements in each direction for these tests and only produce results for the discretisation order $N=3$. As before we run $600$ tests for each case, with and without limiter. As before, the run continued for $200$ time steps. 

The results are given in Table \ref{table:timestep3}. Unlike in the Euler equations, even very small bitflips led to an error above $10^{-14}$. The majority of resulting errors were around $10^{-12}$, so this is not necessarily a sign that SDCs are a considerable  problem in the Einstein equations. Further, note that considerably fewer bitflips led to the program aborting. This is probably due to the relative lower number of variables with physical admissibility criteria attached to them. The first order formulation of the Einstein equations is linearly degenerate and cannot generate shocks, the resulting fields are smooth everywhere except at singularities. As such, in the Einstein equations only one out of the total $59$ variables is constrained, the lapse. The lapse has no physical meaning, however, it can be used to detect how close to a singularity a point is, i.e. how curved space-time is at a point.

\subsection{Soft Errors in the Limiter}

\begin{figure}[t]
  \centerline{\includegraphics[width=0.45\textwidth]{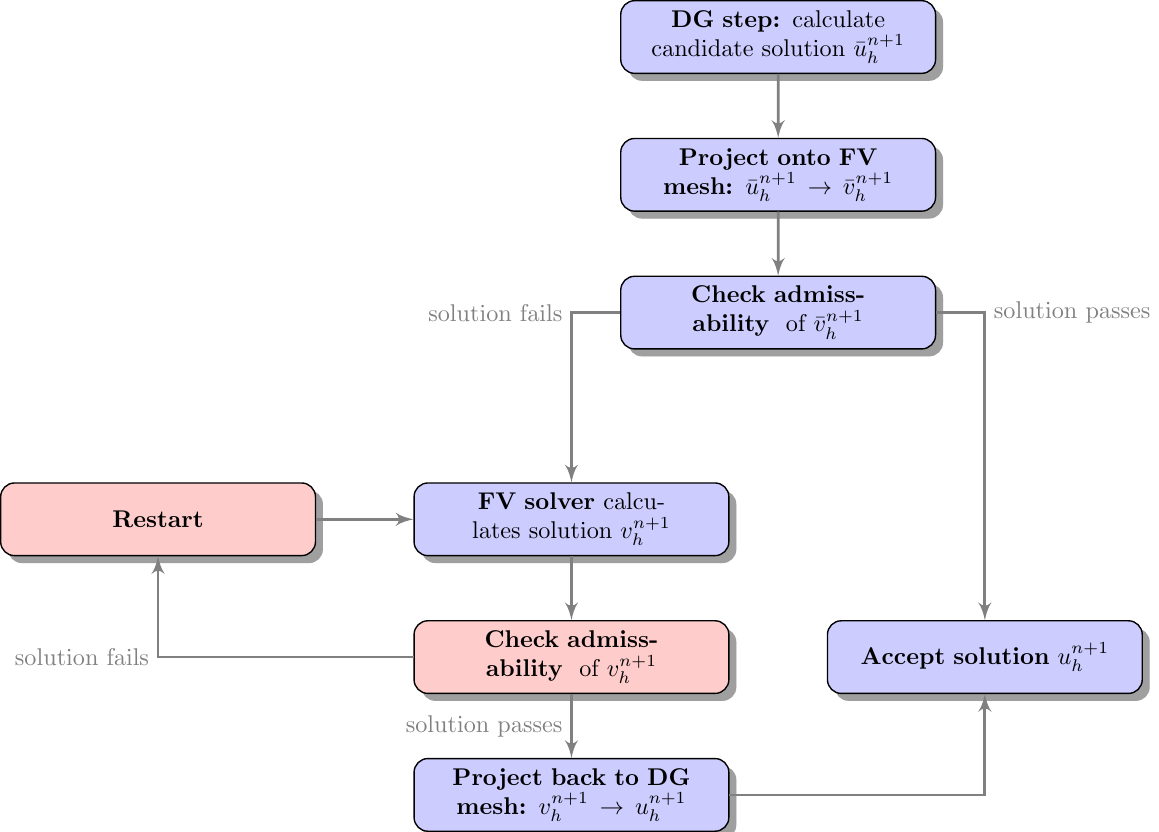}}
 \caption{Schematic of the a-posteriori limiting approach for the time step $n+1$. }
 \label{fig:flowchart_mod}
\end{figure}

In the previous section we examined bitflips which occured during the ADER DG algorithm itself. However, in Section \ref{sec:cost_limiter} we saw that when a limiter is required it takes up a large percentage of the total run time of the program. 
Thus, it seems reasonable to analyse the effect of bitflips in the limiter stage as well. To mitigate errors in this stage we propose a modified limiter step as shown in Figure \ref{fig:flowchart_mod}. After the finite volume solution has been calculated we again check the admissibility criteria and rerun the entire time step if they are not met. 

In this section we use the sod shock tube benchmark example, which requires a limiter to deal with the discontinuities in the initial condition and resulting shocks. Since the limiter makes up a particularly large proportion of the total run time for systems of PDEs with a low number of computed quantities and high polynomial degrees, we test the Euler equations with $N=9$. The test used a mesh of $81$ elements in each direction with two levels of  adaptive refinement allowed near the shock. The test ran for $200$ times steps and was repeated $400$ times for the modified and unmodified algorithm. Instead of adding the bitflips in the space-time predictor step we insert them into the FV scheme during the limiting stage. 

\begin{figure}[b]
  \centerline{
  \includegraphics[width=0.35\textwidth]{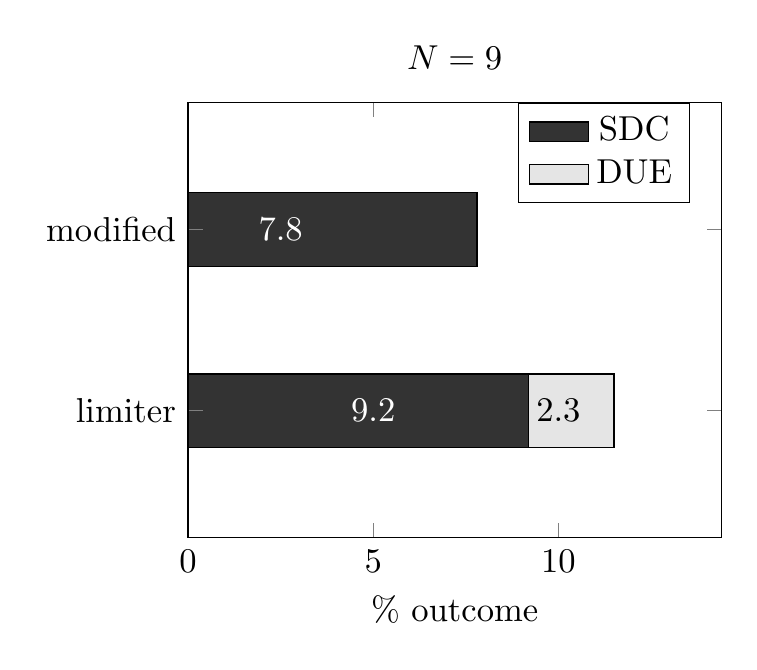}}
 \caption{Effects of a single  bitflip, limiter denotes the original algorithm and modified the modified algorithm with an additional admissibility check. }
 \label{fig:limiter1}
\end{figure}

In Figure \ref{fig:limiter1} and Table \ref{table:timestep_lim} the results of introducing bitflips into the standard limiter-based and modified algorithm are shown. As is to be expected, the results are similar to introducing an error in the algorithm without a limiter for the unmodified algorithm and similar to introducing the error into the limited algorithm for the modified algorithm. In particular, through the second checkpoint we attain results comparable to inserting an error into the space-time predictor in the limited case. The differences in the two test cases are likely a result of the two different initial conditions chosen.

\begin{table}[htb]
\caption{Frequency of outcome and average time step at which the bitflip was inserted into the limiter for different outcomes.
}
\begin{center}
\begin{tabular}{|c|c|c|c|c|}\hline
                & \multicolumn{2}{c|}{limiter} & \multicolumn{2}{c|}{modified}\\
Case            & outcome & time & outcome & time \\\hline
success         & 88.5\% & $104$ &92.2\% & $101$ \\
SDC          & 9.2\%  & $70$  &7.8\% & $89$\\
DUE          & 2.3\%  & $72$  & 0\%  & -\\
\hline
\end{tabular}
\end{center}
\label{table:timestep_lim}
\end{table}

While this procedure is useful in avoiding undetectable errors in the limiter, it comes at a considerable run time cost. The modified algorithm had an average runtime of $132\%$ of the algorithm without the additional checkpoint. Here the average runtime for the unmodified algorithm did not take the runs that did not complete into account. The high cost is a direct result of the fact that checking the admissibility criteria is very expensive.  The local finite volume recalculations themselves are not costly. However, it is important to note that the second admissibility check is only performed on cells on which the limiter is applied. In this test case the limiter is needed in a very large number of cells (in later time steps almost 30\%), in a test case with a low number of cells requiring limiting, the modified algorithm would be considerably cheaper.

\section{Key Findings}

In this paper we examined the inherent resilience of a-posteriori limiting procedures on the fault tolerance of an ADER DG solver. We found that the limiter can indeed increase the resilience of the overall algorithm by detecting faults which would lead to the program aborting. We found that the DUE frequency depends on the hyperbolic system being studied: in our examples, the Einstein equations cannot produce shocks and have a much lower DUE rate than the Euler equations. However, the SDC rate for the Einstein equations is higher, possibly due to insufficiently strong admissibility criteria for the limiter. Stronger admissibility criteria might allow more SDCs to be detected and corrected. Further, we proposed a modification of the algorithm that extends the resilience of the algorithm and showed that it successfully deals with soft faults injected in the limiter step.

The discrete maximum principle's ability to detect large faults is not dependent on the ADER DG algorithm or on the particular a-posteriori limiter described in this paper. We believe that this feature can be exploited more generally in any compute-check-recompute procedure. Similarly, the physical admissability criteria can be applied to check the solution. However, in other numerical schemes, such checks may come at a higher cost. In general, we found that the efficacy of these criteria depends strongly on the PDE system being studied.


\section*{Acknowledgment}
The authors gratefully acknowledge the compute and data resources provided by the Leibniz Supercomputing Centre (www.lrz.de).
Thanks are due to all members of the ExaHyPE consortium who made this research possible.
This work has received funding from the European Union's Horizon 2020 research and
innovation programme under grant agreement No~671698.

\bibliographystyle{plain}
\bibliography{ReportBibD43}

\end{document}